\documentclass[11pt,twoside]{article}

\usepackage{asp2006}
\usepackage{natbib}
\begin{document}
\title{Radio observations of colliding winds in massive stars}   
\author{Sean M. Dougherty}   
\affil{National Research Council of Canada, Herzberg Institute of Astrophysics}    

\begin{abstract}
This brief review describes radio observations of colliding winds in
massive stars starting with the first direct observational support for
the colliding-wind model advanced in the early 1990's to explain
non-thermal radio and thermal X-ray emission in some massive
stars. Studies of the well-studied and highly-eccentric WR+O star
system WR\,140 are described along with recent observations of O-star
systems. Also discussed is the binary nature of almost all massive
stars that exhibit non-thermal behavior and some strategies for
finding new systems.
\end{abstract}


\section{Introduction}
Massive stars have dense stellar winds that are photo-ionized by the
strong UV radiation fields from the underlying star, giving rise to an
excess of free-free continuum emission over the photospheric emission,
that can be observed from IR to radio wavelengths.  The brightness
temperature of this emission is $\sim10^4$~K, as expected from a
photo-ionized envelope. The emission is partially optically thick, and
between mid-IR and radio frequencies has a power-law spectrum
$S_\nu\propto\nu^\alpha$, with spectral index ($\alpha$) typically
$\sim+0.7$ \citep[e.g.][]{Williams:1996}.  A number of massive stars
have radio emission properties that differ from this typical picture:
they exhibit brightness temperatures $\sim10^6-10^7$~K, and flat or
negative spectral indices, both properties characteristic of
non-thermal synchrotron emission and therefore of relativistic
electrons. Often, this non-thermal emission is variable
\citep{Abbott:1984, Abbott:1986}.

It was first noted by \cite{vanderhucht:1992} that many Wolf-Rayet
(WR) stars exhibiting non-thermal emission were long-period binary
systems, and they proposed that the non-thermal emission arose from the
region where the massive stellar winds of the two binary components
collided, the wind-collision region (WCR) outside the optically thick
part of the circumbinary stellar wind envelope. The colliding-wind
binary (CWB) model was described in the seminal paper of
\cite{Eichler:1993}.

Spatially resolved observations of the WR+O system WR~147
\citep{Williams:1997} presented the first unequivocal confirmation of
the CWB model. The model is supported further by the dramatic
variations of the synchrotron radio emission in the 7.9-year WR+O
binary system WR~140, that are clearly modulated by the binary orbit
(c.f. Fig.~\ref{fig:wr140data}).  Among WR stars, the CWB model has
now been substantiated with over 90\% of WR stars that exhibit
non-thermal emission are either a binary system or have a ``nearby''
massive stellar companion that results in a WCR
\citep{Dougherty:2000a}.  Amongst O-type stars, there is now
increasing evidence that a similar explanation also holds (see papers
by van Loo, and Benaglia in these proceedings;
\cite{debecker:2007}). The nature of the acceleration mechanism is, as
yet, unidentified but is widely attributed to diffuse shock
acceleration in shocks that ``bound'' the WCR. Alternative mechanisms,
such have magnetic reconnection have been discussed
\citep[e.g.][]{Jardine:1996}.

Colliding-wind binary systems present an important laboratory for
investigating the underlying physics of particle acceleration, of not
only the relativistic electrons that give rise to the observed radio
emission, but potentially to ion acceleration that may give rise to
gamma-ray GeV-TeV emission (see other papers in these proceedings). In
addition to the excellent modelling constraints provided from
high-resolution VLBI imaging, radiometry and most recently high-energy
observations, CWBs provide access to higher mass, radiation and
magnetic field densities than found in supernova remnants, that have
been widely and successfully used for studying particle acceleration,
particularly diffusive shock acceleration (see Ellison, these
proceedings).

Here the observational evidence for non-thermal radio emission arising
in WCRs in both WR and O-star binary systems is reviewed. In closing,
some potential schemes for finding more of these particle acceleration
laboratories are suggested.

\section{Observational evidence for colliding-wind systems}

\subsection{The first essential proof of a CWB - imaging WR\,147}

\begin{figure*}
\plottwo{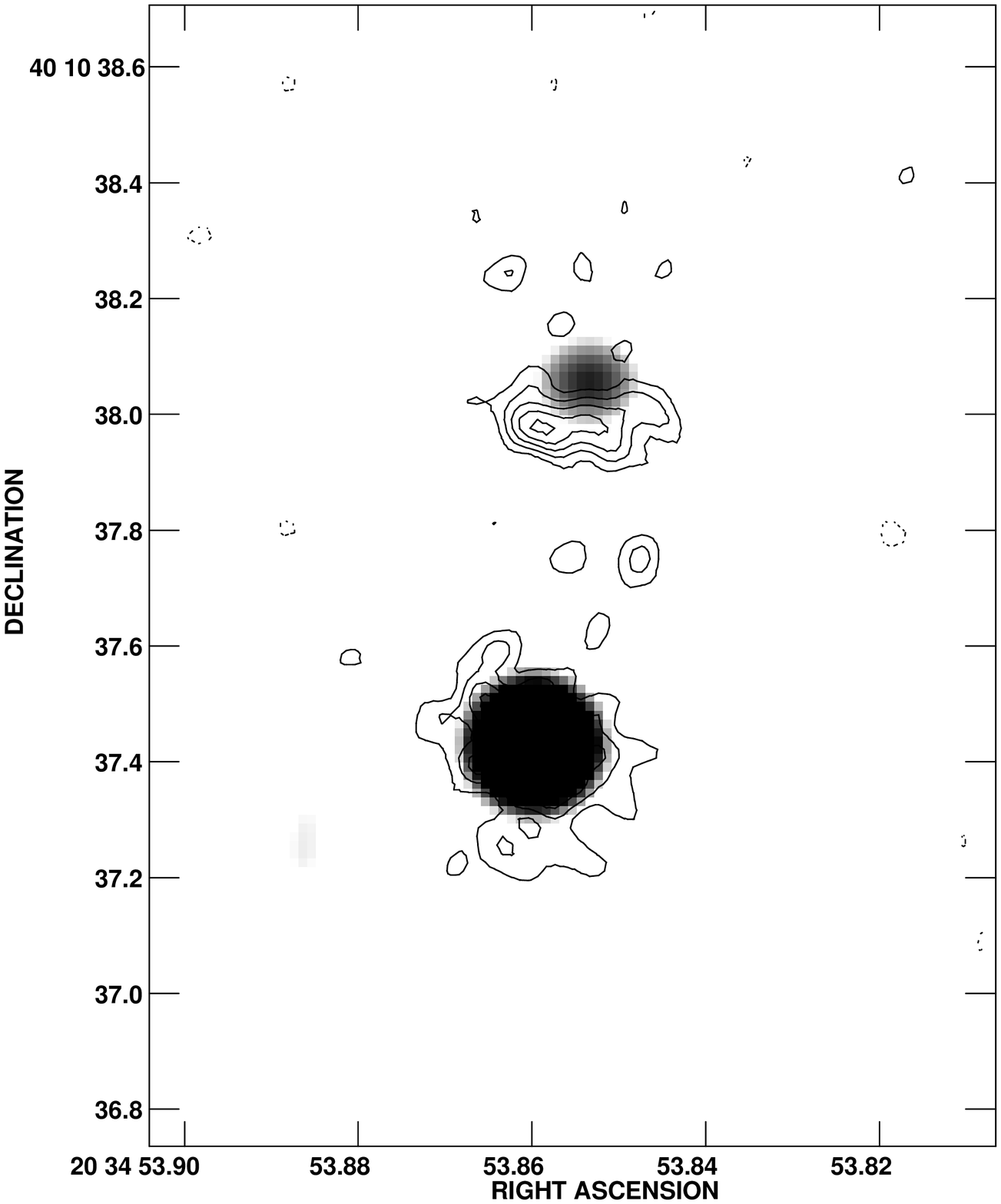}{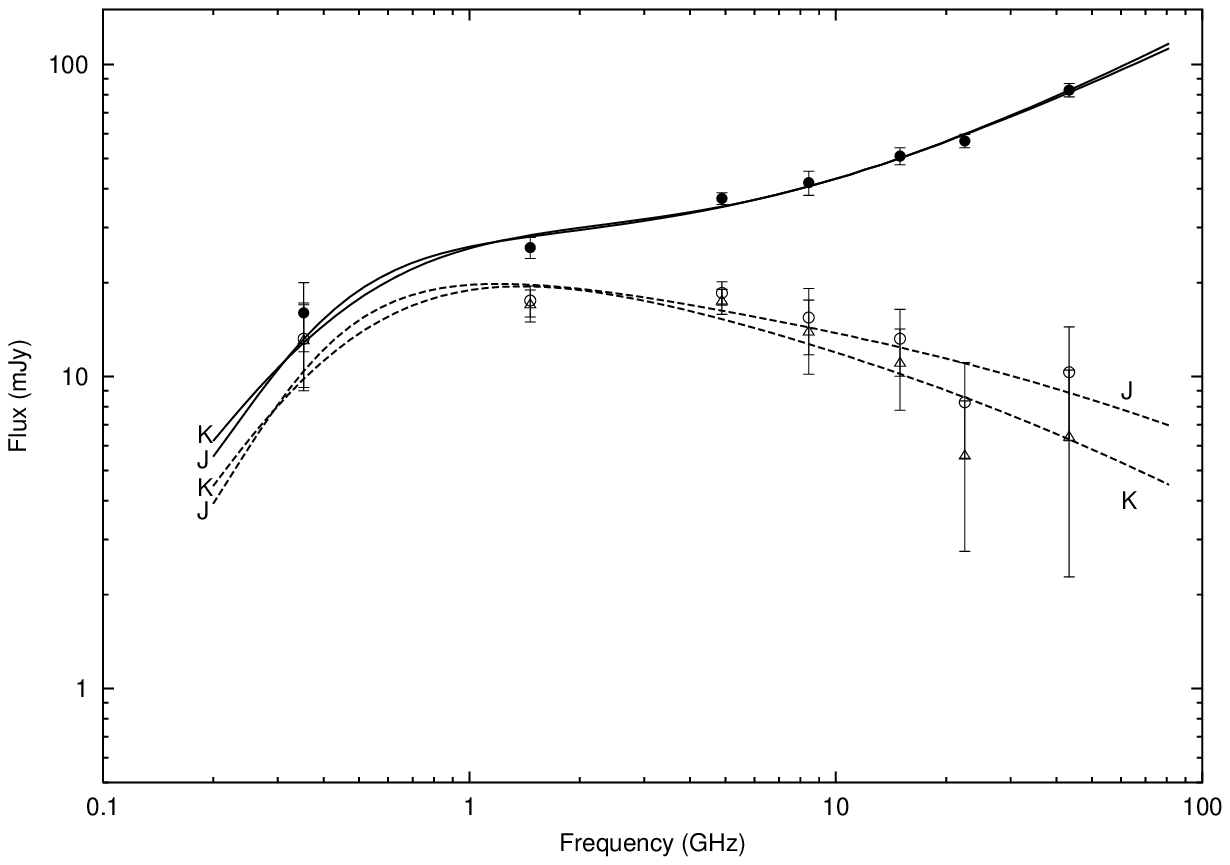}
\caption[]{Left: Overlay of a MERLIN 5-GHz image of WR147 (contours)
and a $2.2\mu$m UKIRT image (greyscale). The IR image of the WR star
was aligned with the peak position of the thermal emission in the
southern radio component, assumed to be the ionized stellar wind of
the WR star. The position of the non-thermal emission region (the
northern radio component) is consistent with the location of ram
pressure balance between the winds of the WR star and the B-type
companion revealed in the IR image \citep{Williams:1997}. Right: Radio
fluxes from WR147 with model fits to both the total (solid symbols)
and the non-thermal emission component (open symbols) based on
2D-hydrodynamic models of the WCR and stellar winds
\citep[from][]{Pittard:2006a}. Note the large uncertainties in the
fluxes, particularly at the higher frequencies.}
\label{fig:wr147}
\end{figure*}

\cite{Williams:1997} provided the first direct imaging confirmation of
the colliding-wind model in observations of WR\,147. This system was
shown by \cite{Moran:1989} using the MERLIN telescope array in the UK
to have two distinct components; a thermal component associated with
the optical position of the WR star, and a non-thermal component 0.6
arcseconds to the North. \cite{Moran:1989} speculated that this may be
due to the presence of a companion star as had been suggested
previously to explain non-thermal emission in WR140
\citep{Williams:1987}.  No companion could be found in a search for
objects down to 1.9 magnitudes fainter than the WR star in optical
R-band observations.  Soon thereafter, \cite{Churchwell:1992} made
multi-wavelength radio observations with the Very Large Array (VLA)
and confirmed the nature of the emission components, but also did not
identify a companion star. A compact companion was suggested to
explain the X-ray and non-thermal radio emission.  It was not until
\cite{Williams:1997} carried out sufficiently high-resolution imaging
at 2.2 microns with UKIRT that a nearby companion was identified, 3
magnitudes fainter than the WR star at 2.2 microns and with the
luminosity of a B0-type star. The companion is 60 mas to the north of
the non-thermal radio component, in a position consistent with the
stellar wind momentum ratio of a WN8 and a B0 star (see
Fig.~\ref{fig:wr147}) and with enough kinetic luminosity to power the
X-ray emission.

Subsequently, multi-frequency radio data of WR\,147 was used to
attempt to constrain models of the emission from CWBs
\citep[e.g.][]{Skinner:1999, Dougherty:2003}. However, because the
thermal emission from the stellar wind of the WN8 star in WR\,147
dominates at radio wavelengths, the continuum spectrum of the
non-thermal emission is not of sufficient signal-to-noise to constrain
the models well. Better model constraints have had to wait for the
observations of other systems.

\begin{figure*}[t]
\plottwo{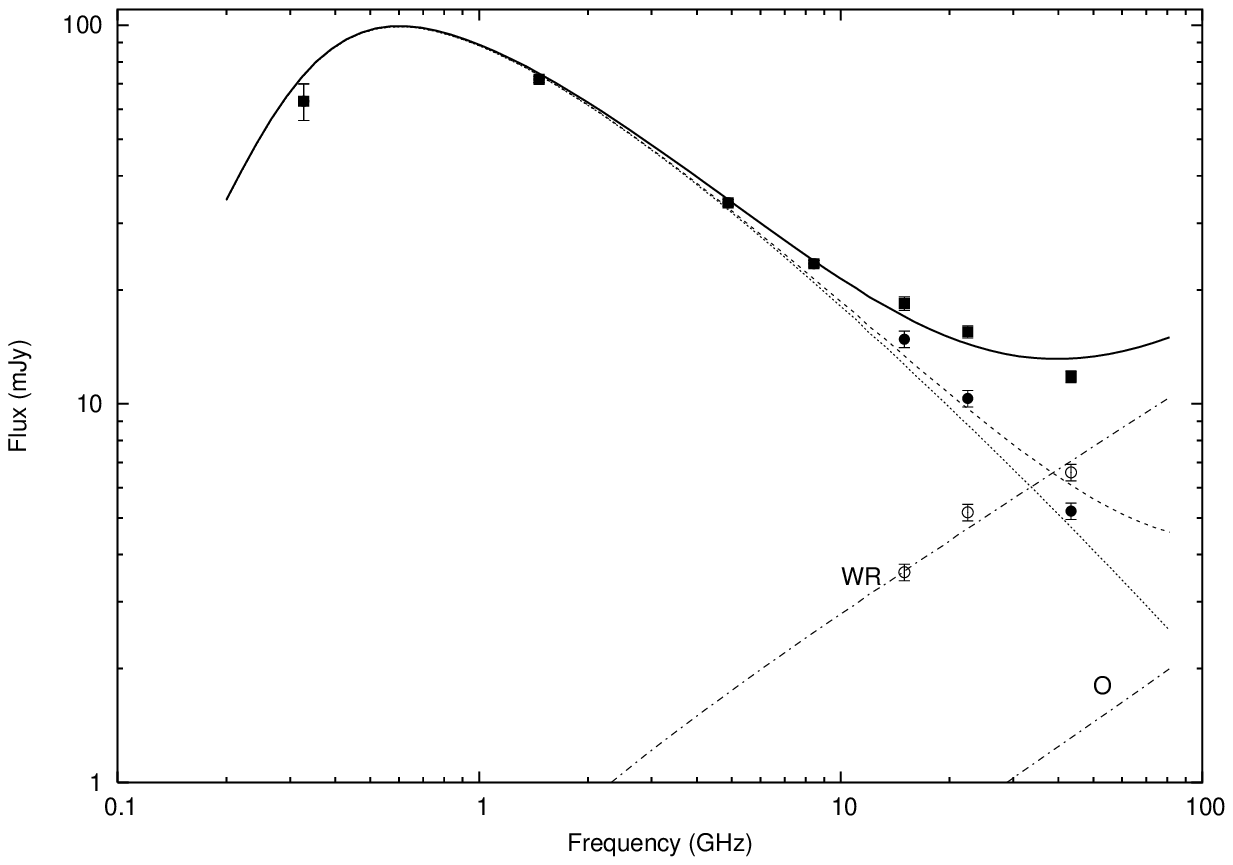}{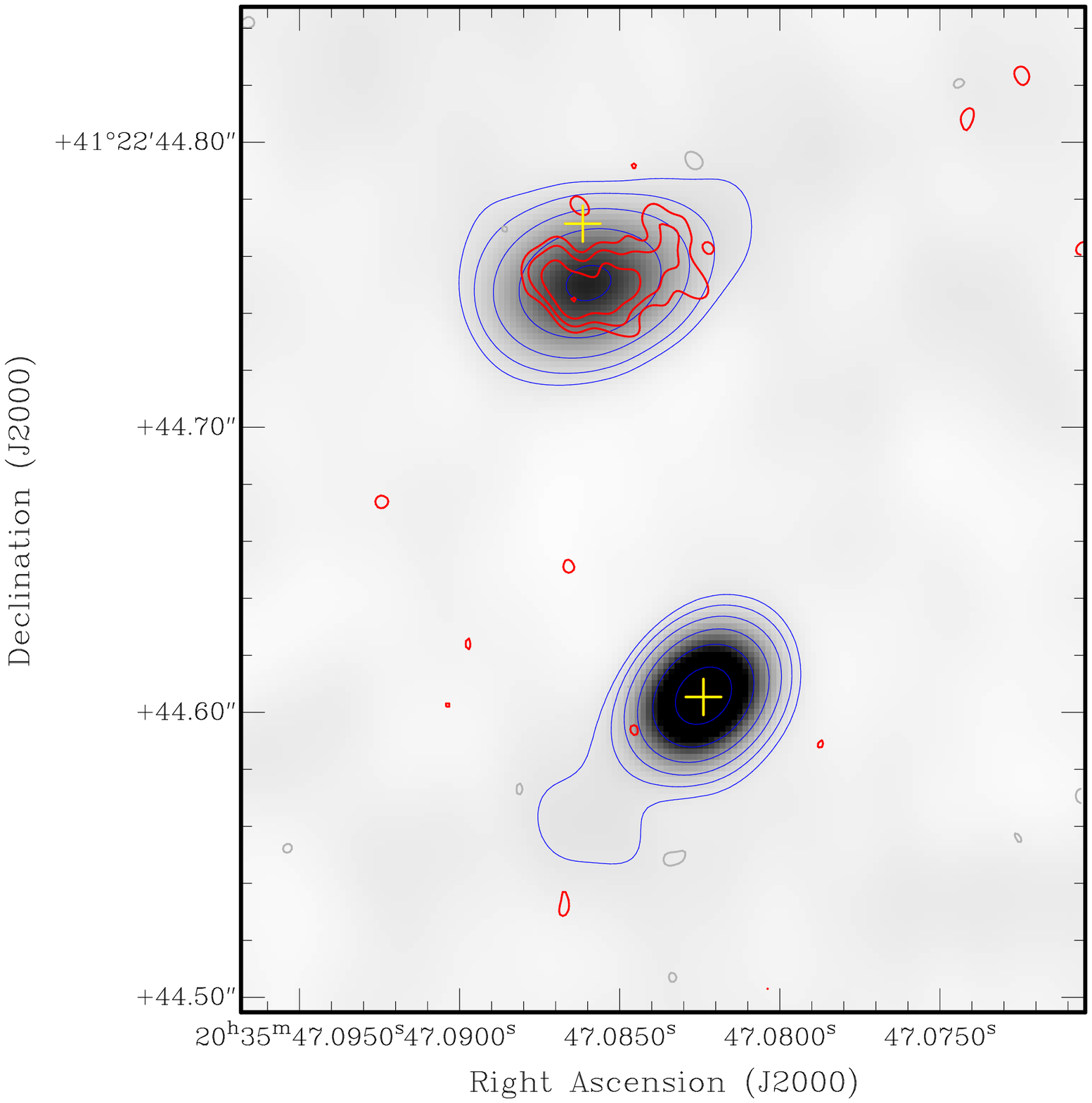}
\caption[]{Left: Best-fit models of the radio emission from the WCR in
WR146, with synchrotron (dotted), thermal (lower dot-dashed) and
synchrotron+thermal (dashed). The thermal emission from the WR star
wind is the higher dot-dashed line.  The solid line is the total
flux. Right: High-resolution images of WR146 from VLA 43-GHz (blue)
and VLBA 8.6-GHz (red) observations. The radio observations are
aligned after accounting for proper motion between the two
observations. The 43-GHz observations show the southern thermal source
that is associated with the WR star and the northern non-thermal
emission from the WCR. The VLBA observation reveals the structure of the
WCR. The crosses mark the relative location of the two stars, assuming
the WR star is coincident with the southern 43-GHz thermal source
(O'Connor et al., in preparation).}
\label{fig:wr146}
\end{figure*}

\subsubsection{WR\,146 - toward better modelling constraints}
At a similar time as the WR\,147 observations,
\cite{Dougherty:1996} found evidence in MERLIN 5-GHz observations for
a wide colliding-wind binary system in WR\,146. This assertion was
supported by observations from the HST that revealed an O-type star
168 milli-arcsecond north of the WC star \citep{Niemela:1998}. Unlike
WR\,147, the radio emission from this system is dominated by the
non-thermal emission arising from the WCR and so is a more ideal object
for both very high resolution multi-wavelength imaging and high-quality
radiometry that give more robust modelling constraints. Initial models
have been constructed for this system \citep{OConnor:2005} (see
Fig~\ref{fig:wr146}).

\section{WR\,140 - the particle acceleration laboratory}
Both WR\,146 and WR\,147 are ``visual'' binary systems. As such, the
WCR occurs far out in the stellar wind of the dense WR star, and the
lines-of-sight to the WCR are typically optically thin.  The apparent
separation of the stars ($a \sin i$, where $a$ and $i$ are the orbit
size and inclination respectively) in these systems suggest that, with
an assumed binary mass of $\sim50$M$_\odot$, the orbital periods are a
minimum of $\sim350$~yrs. In such systems, variations due to the
relative motion of the stars will all have long timescales - long
relative to the time frame of typical study! Such systems lack a
well-defined ``orbit'', most particularly the separation of the
stars. In spite of the apparent success modelling the continuum
spectra and spatial distribution of the radio emission in both WR\,146
and WR\,147 \citep[see][]{Pittard:2006a, OConnor:2005}, this presents
an impediment to modelling as a number of key system criteria remain
ill-defined e.g. the UV-radiation density and the wind-kinetic
luminosity available within the WCR.

Closer systems, with periods of a few years and orbits that are
well constrained through spectroscopic studies, yet are sufficiently
wide for optically thin lines-of-sight to the WCR through the
circumbinary stellar wind plasma are better suited for modelling. In
addition to excellent system constraints, those with particularly
eccentric orbits have variable conditions within the WCR that offer an
additional probe of the underlying physics.

\begin{figure*}
\begin{center}
\plottwo{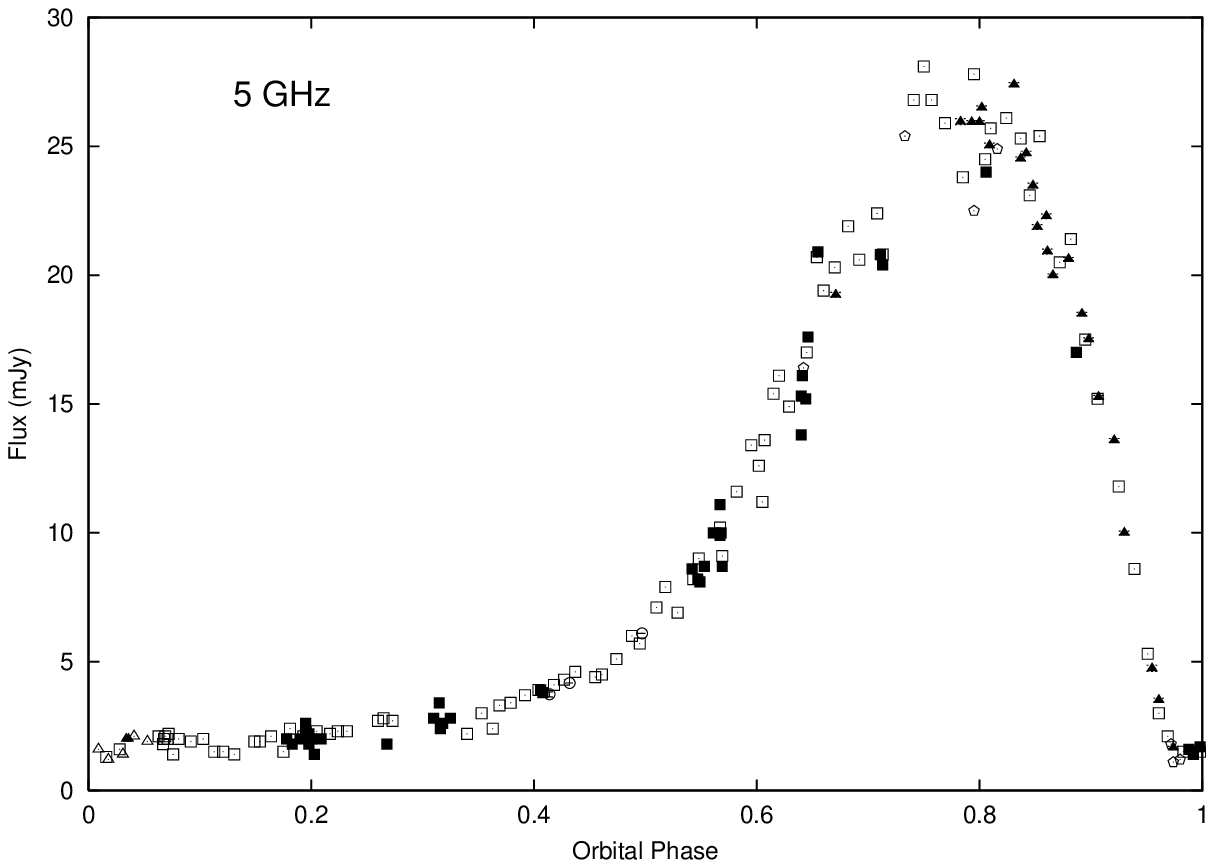}{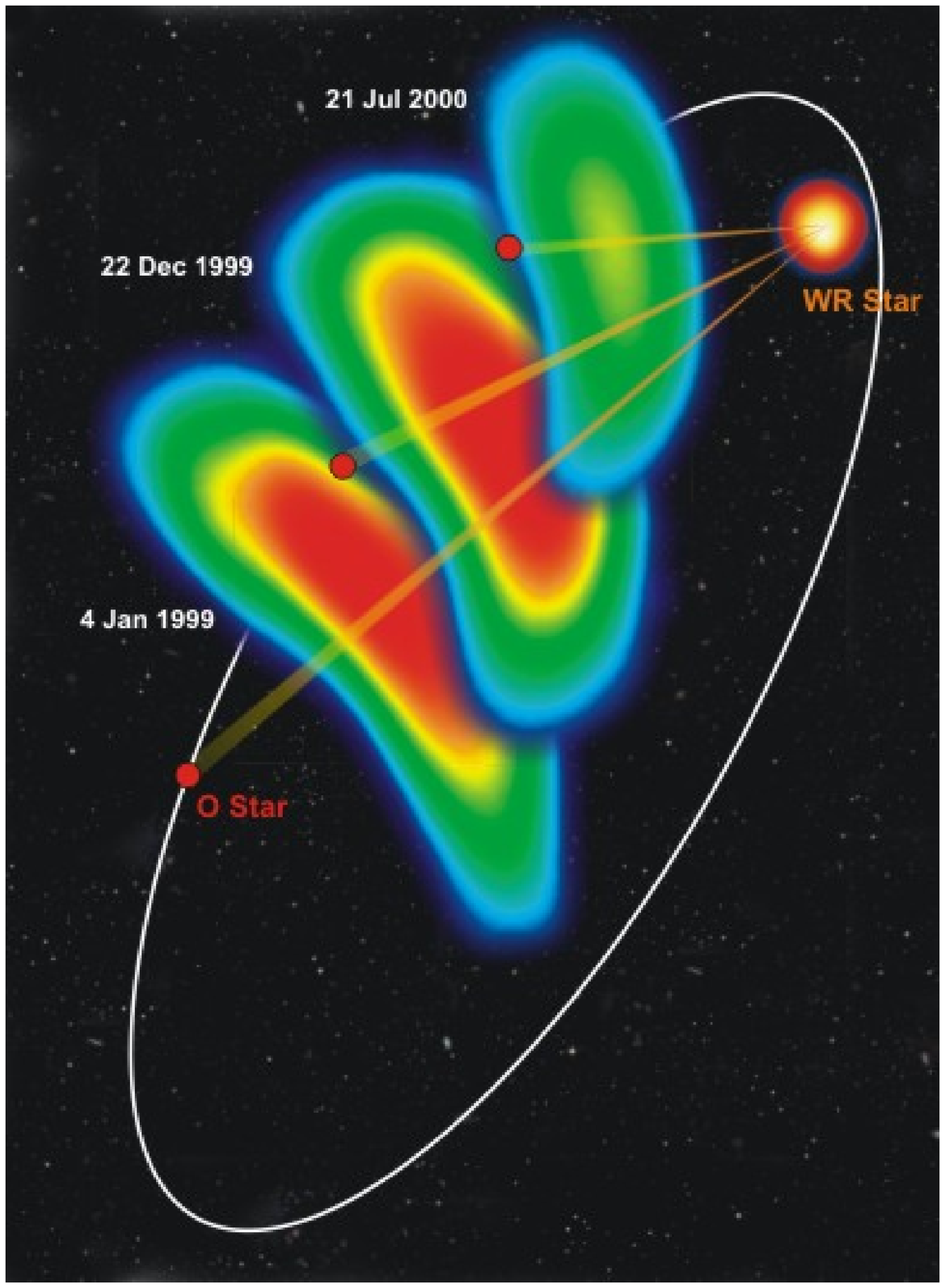}
\end{center}
\caption[]{Left: 5-GHz VLA radiometry of WR\thinspace140 as a function
 of orbital phase for orbit cycles between 1978-1985 (pentagons),
 1985-1993 (squares), 1993-2001 (triangles), and the current cycle
 2001-2009 (circles). Open symbols are from the VLA \citep{White:1995,
 Dougherty:2005} and solid symbols from the WSRT \citep{
 Williams:1990, Williams:1994}. Right: A montage of 8.4-GHz VLBA
 observations of WR\thinspace140 at three different orbital phases
 that demonstrate the rotation of the WCR as the orbit progresses. The
 deduced orbit is superimposed \cite[][]{Dougherty:2005}.
\label{fig:wr140data}}
\end{figure*}

The 7.9-yr period, highly eccentric WC7+O4 binary WR\,140 is one such
system, and is the archetype CWB system. It is well known for its
dramatic variations in emission from near-IR to radio wavelengths
\citep{Williams:1990, White:1995}, and also at X-ray energies 
throughout its orbit \citep{Zhekov:2000,Pollock:2002,Pollock:2005}.
The WCR in WR140 experiences significant changes in its local
environment as the stellar separation varies between $\sim 2$~AU at
periastron and $\sim 30$~AU at apastron.  The observed radio emission
increases by up to two orders of magnitude between periastron and a
frequency-dependent peak between orbital phases 0.65 and 0.85,
followed by a steep decline. This behavior repeats almost exactly from
one orbit cycle to another, suggesting a particle acceleration
mechanism that is well-controlled by the orbit of the system
(Fig.~\ref{fig:wr140data}).

\subsection{The first fully defined CWB orbit geometry}

To fully establish the geometry of the WR140 binary system,
particularly the orbital inclination ($i$) and semi-major axis ($a$),
but also the longitude of the ascending node ($\Omega$), the system
must be resolved into a ``visual'' binary. This was achieved by
\cite{Monnier:2004} using the Infrared-Optical Telescope Array (IOTA)
interferometer. Together with other established orbit parameters
\citep[e.g.][]{Marchenko:2003}, families of possible solutions for
($i,\Omega,a$) can be determined. Until similar observation at other
orbital phases are available, VLBI observations of the WCR are the
only means to determine uniquely $i$, and hence $\Omega$ and $a$, from
these solution families.

A 24-epoch campaign of VLBA observations of WR140 was carried out
between orbital phase 0.7 and 0.9 \citep{Dougherty:2005}. At each
epoch, an arc of emission is observed, resembling the bow-shaped
morphology expected for a WCR. This arc rotates from ``pointing'' NW
to W as the orbit progresses.  The WCR emission is expected to wrap
around the star with the lower wind momentum - the O star.  In this
case, the rotation of the WCR as the orbit progresses implies that the
O star moves from the SE to close to due E of the WR star over the
period of the VLBA observations. Thus, the direction of the orbit is
established.  Secondly, the orbital inclination can be derived from
the change in the orientation of the WCR with orbital phase.  Each
($i,\Omega$) family provides a unique set of position angles for the
projected line-of-centres as a function of orbital phase. This gives
$i=122\deg\pm5\deg$ and $\Omega=353\deg\pm3\deg$, leading to a
semi-major axis of $a=9.0\pm0.5$~mas, and the first full definition of
orbit parameters for {\em any} CWB system.  From optical spectroscopy
$a\,\sin i = 14.10\pm0.54$~AU \citep{Marchenko:2003}, implying a
distance to WR\,140 of $1.85\pm0.16$~kpc. Notably, this estimate is
independent of a calibration of stellar parameters e.g. absolute
magnitude, and implies that the luminosity of the O star is that of a
supergiant.

\begin{figure*}
\plotone{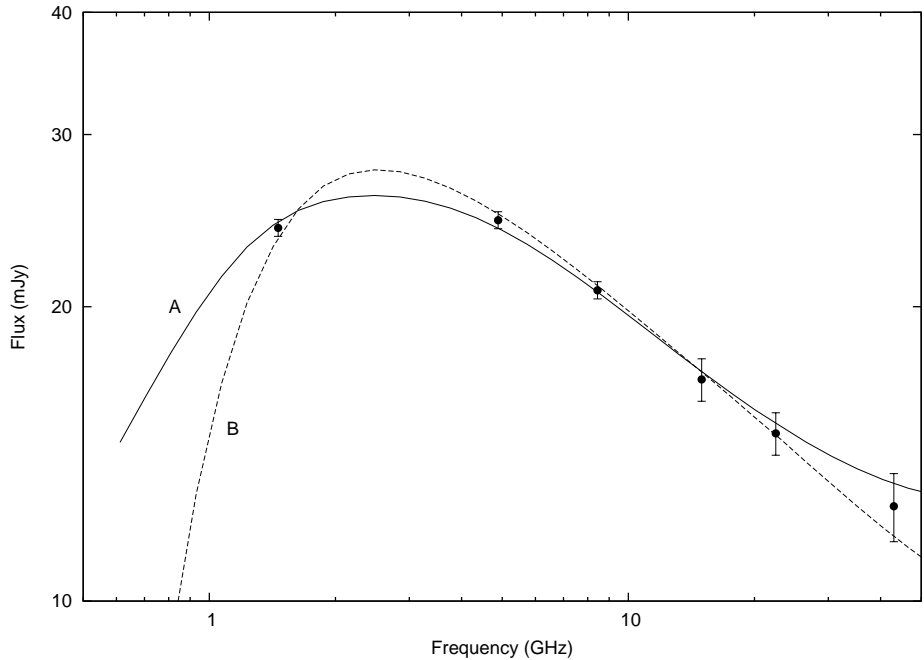}
\caption[]{A couple of models of the radio emission from WR\thinspace140 at
orbital phase 0.837. Note the importance of a high-precision 43-GHz
datum to distinguishing models. \label{fig:wr140models}}
\end{figure*}

\subsection{Modelling the radio emission from WR140}
With the geometry and stellar parameters, models of the WCR in WR\,140
have better observational constraints than in any other known CWB. In
addition, observations of the thermal X-ray emission that arises from
the shocked plasma in the WCR provide the best available constraints
on the mass-loss rates of the two winds as a function of the momentum
ratio of the winds. 

\cite{Pittard:2006b} studied models of the radio and X-ray emission at
orbital phase 0.837, where non-thermal emission dominates the radio
spectrum and a good estimate of the characteristics of the
relativistic electrons is possible.  Fig.~\ref{fig:wr140models} shows
two model fits to the radiometry, with quite different values of
relative wind-momentum.  The models suggest the low-frequency turnover
at $\sim 3$~GHz is due to free-free absorption in the winds, since
models where the Razin effect dominates the low-frequency spectrum
place an unacceptably large fraction of the shock energy in
non-thermal electrons. Development of these models continues (see
Pittard, these proceedings).

Observations over a wide radio frequency range, and particularly those
at the higher and lower frequencies, are key to distinguishing between
models (see Fig.~\ref{fig:wr140models}). Currently, 
radiometry is the only available technique since high-resolution radio
observations made to date do not have sufficient low brightness
sensitivity combined with resolution to detect lower brightness
emission far out in the downstream flow of the WCR that may reveal the
asymptotic opening angle of the WCR, a function of wind-momentum
ratio. The current upgrade to the Very Large Array (the EVLA) will
significantly improve radiometry precision, and the broad-banding of
VLBI networks e.g. VLBA, eMERLIN will improve high-resolution imaging
capability, most especially when combined with the data from the EVLA.

\begin{figure*}
\plottwo{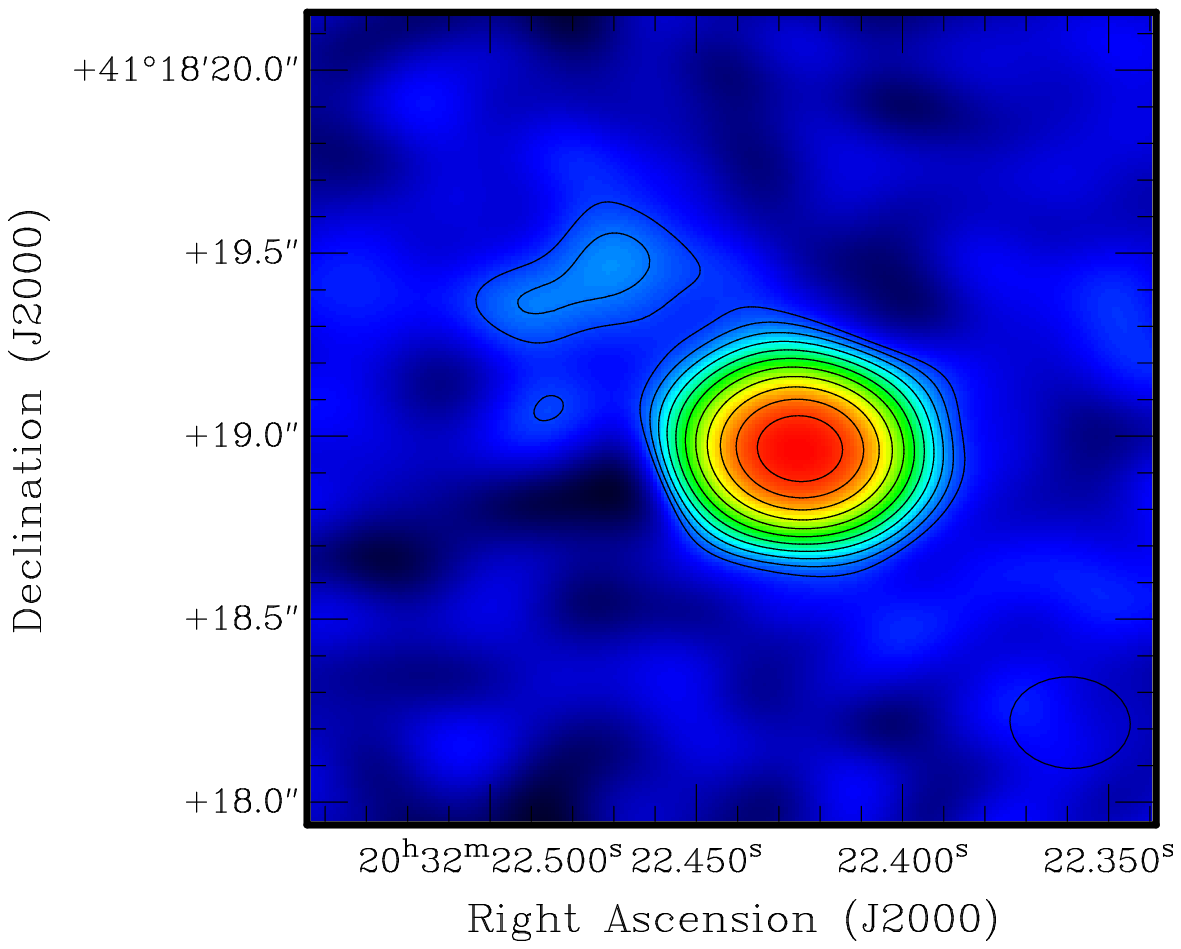}{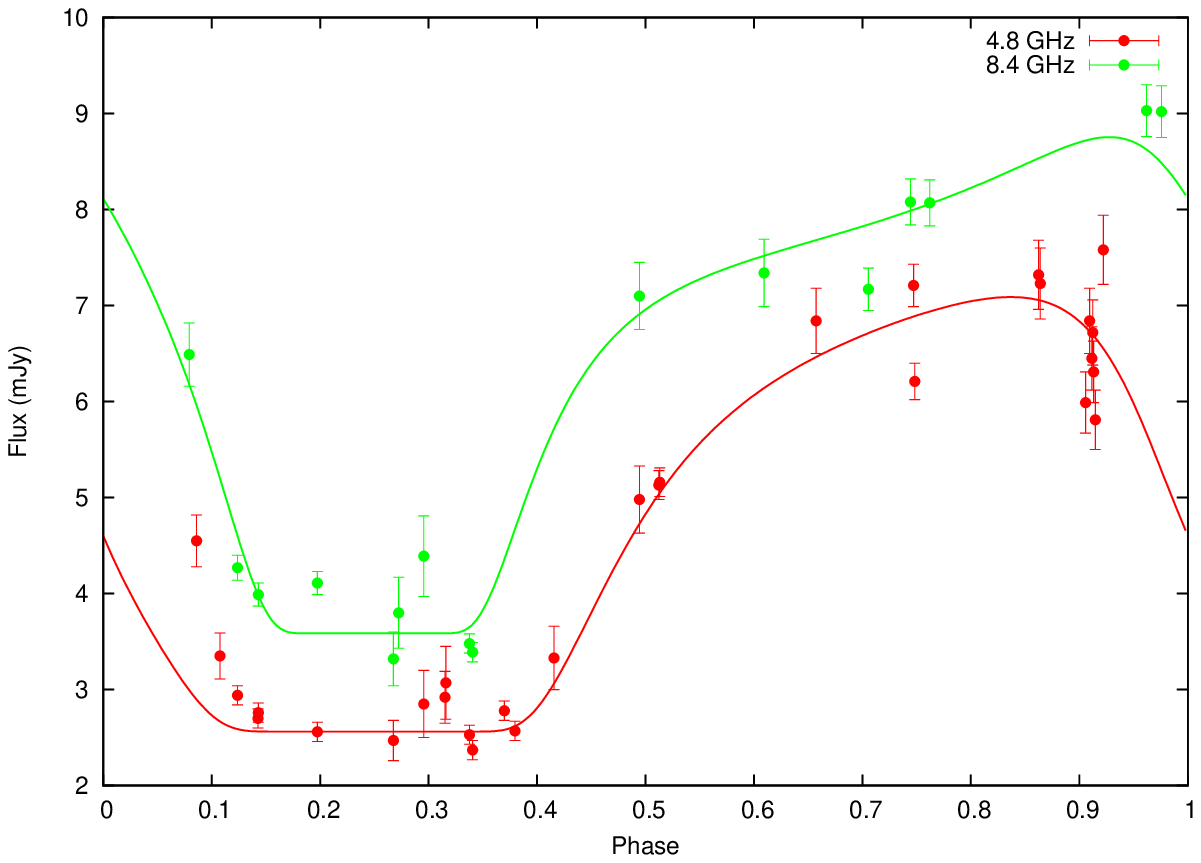}
\caption[]{Left: An 8.4-GHz image of Cyg OB2 \#5 from the VLA, that
clearly shows two components. The bright component is ``thermal'' and
associated with the 06-binary system, and the weak component to the NE
is non-thermal.  Right: The radiometry at 4.8 and 8.4 GHz of the
bright component, folded with a 6.7-yr orbit period and showing the
best-fits at the two frequencies of a simple free-free opacity model
with a non-thermal source in a 6.7-yr eccentric orbit around the
binary (Kennedy et al, in preparation).
\label{fig:cygob25}}
\end{figure*}

\section{Looks like a duck, quacks like a duck - what is it? O-star binaries}

The bulk of observations and modelling of CWB systems has been
established using radio-bright WR+OB binary systems.  Given that these
are the descendents of massive O-star binaries, it begs the question
of whether the CWB model can also explain non-thermal emission amongst
O stars.  Certainly, of the 16 O-star systems that exhibit non-thermal
emission ($\sim40\%$ of all the radio-detected O stars), at least
$70\%$ are established binary systems \citep{debecker:2007}. Here, a
couple of systems with good observational evidence for a WCR are given as
examples.

\begin{figure*}
\plottwo{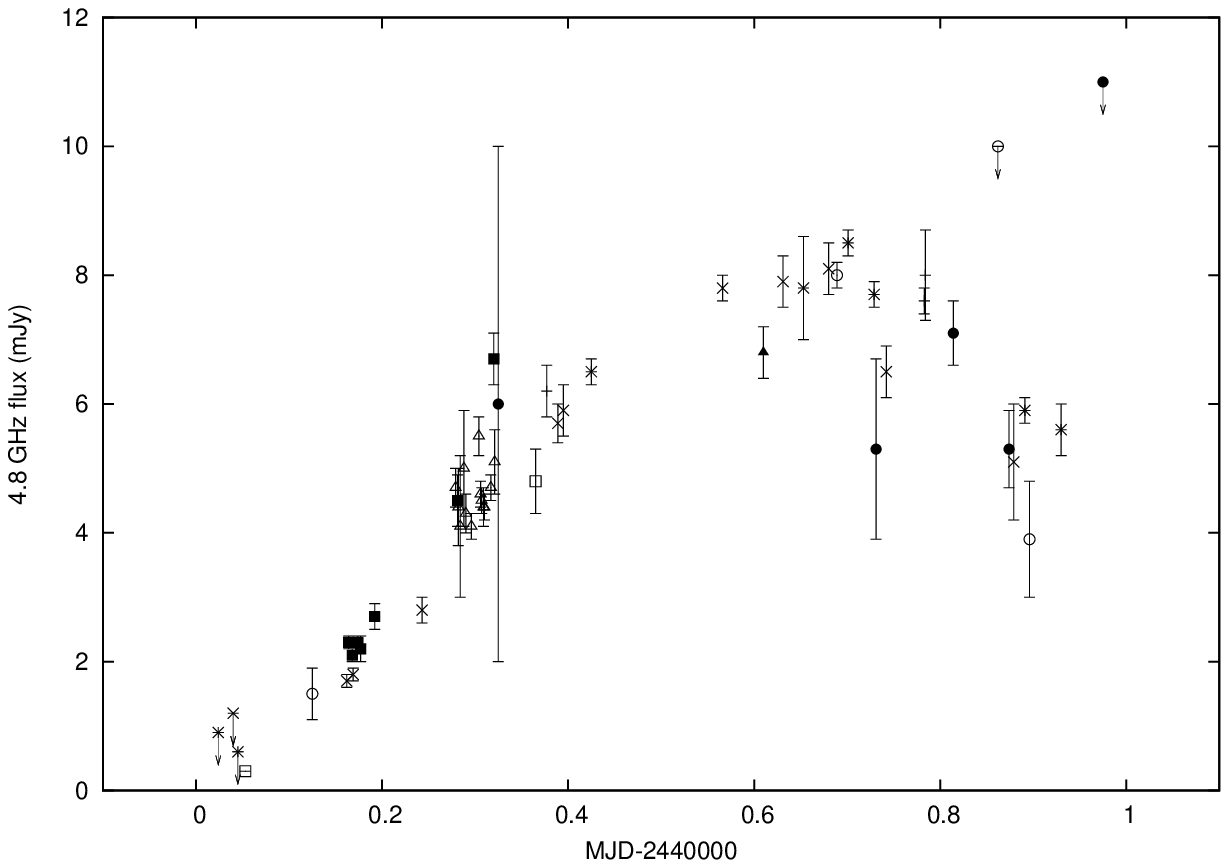}{dougherty_fig6b.eps}
\caption[]{Left: 5-GHz flux of Cyg OB2 \#9, modulated by a 2.35-year
period, based on a preliminary reduction of archive VLA data.  Phase
0.0 was arbitrarily set at radio minimum. Right: An 8.4-GHz VLBA
observation of Cyg OB2 \#9.  A bow-shaped WCR is clearly resolved
(Dougherty et al., in preparation)
\label{fig:cygob29}}
\end{figure*}

In the 6.6-day binary O6+O6 star system Cyg OB2 \#5, is both a thermal
and a non-thermal radio component \citep[e.g.][]{Miralles:1994},
similar to those observed in WR\,146 and WR\,147
(Fig.~\ref{fig:cygob25}). \cite{Contreras:1997} suggested that the
non-thermal component is from a WCR between the O-star binary and a
B-type star $\sim0.9$ arcseconds distant. However, most recently
Kennedy et al.~\citetext{in preparation} have demonstrated that radio
emission from the putative WCR is constant and the radio variations
first noted by \cite{Miralles:1994} arise solely from the ``thermal''
component that is associated with the O6 binary. They succeed in
modelling the variations through orbit-modulated free-free opacity
along lines of sight to a non-thermal source in a 6.7-yr orbit around
the O6-star binary, where at minimum the spectral index is closely
that of a ``classic'' stellar wind plasma ($\sim0.6$), and at maximum
is flatter ($\sim0.25$) due to the non-thermal contribution to the
total emission (Fig.~\ref{fig:cygob25}). Recently \cite{Linder:2009}
suggest that this model may also explain the X-ray emission.  The
success of this simple model is compelling, though multi-frequency
observations may reveal a more complex story - as discovered for
WR\,140!

In Cyg OB2 \#9, analysis of the radio emission from the system shows
that the emission is clearly modulated with a 2.35-year period, very
similar to the behavior of the radio emission from WR\,140. VLBA
observations of this system reveal a bow-shaped non-thermal emitting
region, as anticipated for a wind-collision region
(Fig.~\ref{fig:cygob29}). Together, this evidence suggests strongly a
WCR origin for the emission, and recently \cite{Naze:2008} succeeded
in the long-standing challenge of finding spectroscopic evidence of
the companion required to give rise to the WCR.

The observational support for the CWB model among O-type stars is
rising (see van Loo, these proceedings). The evidence from imaging
and/or radiometry in these examples of recent radio work is remarkably
similar to those of the established WR+OB CWB systems - even in the
absence of supporting evidence for a stellar companion. The evidence
of WCRs in non-thermal emitting O stars continues to grow in new
observations, particularly at higher resolution e.g. HD93129A
\citep{Benaglia:2009}.

\section{Finding new systems}

Identifying stellar companions through the traditional technique of
detecting radial velocity variations can be extremely challenging. It
is strongly dependent on the luminosity (and mass) contrast of the
two stars in the binary systems, and the period and orientation of the
orbit. Certainly for many of the currently known CWB systems, the
periods are long ($>1yr$) and require long-term dedicated observing
programmes over many years to identify \citep[see][for
examples]{Williams:2002}.

Identifying CWB systems is important in a number of leading areas of
astrophysics.  Establishing the binary fraction among massive stars is
crucial for models of binary star formation, especially in young massive
stellar clusters where the long-period binary systems may be readily
disrupted. Secondly, the ability of the WC-type WR binaries to produce
quite large amounts of dusts \citep[e.g.][]{Tuthill:1999} leads to the
question of whether massive binary systems could have provided significant dust
production during the era of first stars and galaxy formation.

The CWB systems that we know of today have been identified through at
least one, and often two of the following: 1) non-thermal radio
emission, 2) thermal X-ray emission or, 3) variable dust emission, all
emission from within the WCR. A number of methods that take advantage of
these observational attributes have been successfully used recently
and suggest methods for identifying binary systems that will not be
readily identified using spectroscopic techniques. Furthermore, these
methods mitigate against the need for long-term monitoring to detect binaries.

\cite{Clark:2008} studied the X-ray properties of the young massive
cluster Westerlund 1 and showed that a number of members have hard
($kT\ge2.0$~keV) and bright ($L_x\sim10^{32-33}$~erg~s$^{-1}$) X-ray
emission as anticipated from a WCR \citep[e.g.][]{Stevens:1992}.
Interestingly, these stars also exhibit either radio continuum spectra
that are too flat to be thermal emission alone
\cite[e.g.][]{Dougherty:2008}, and/or they exhibit an IR excess that
has been attributed to hot dust \citep{Crowther:2006}.  These
identifications have contributed to determining a very high binary
fraction ($>70\%$) amongst the WR stars in Westerlund 1
\citep{Clark:2008}.

Perhaps most dramatically, \cite{Tuthill:2006} used
diffraction-limited IR imaging to readily identify the five `cocoon'
stars in the massive young Quintuplet Cluster as WR-binary systems,
with each having a characteristic CWB dusty ``pinwheel'' nebula. The
difficulty of previous attempts at identifying the character of these
stars, demonstrates the power of this high-resolution technique,
certainly for identifying dusty WC-type WR stars.

Most recently, the ChIcAGO project has been identifying ``anomalous''
ASCA X-ray sources with fluxes in the range
$10^{-11}-10^{-13}$~erg~s$^{-1}$ in the Galactic Plane. Such sources
are not bright enough to be well-known, yet too bright to be canonical
active stars or AGN. Through follow-up observations with CHANDRA (for
resolution), 2MASS and spectroscopy (for stellar identification) and radio
observations a number of these objects have now been identified as
colliding-wind systems (G.~Anderson \& B.~Gaensler, priv. communication).

\section{The future of this research}

High-resolution imaging at radio wavelengths has provided dramatic
confirmation that relativistic electrons are being accelerated in
wind-collision regions in massive binary systems. The evolution of the
WCR revealed in WR\,140 shows what can be achieved with this
technique. However, high quality radiometry is still central to models
of the radio emission, and other high-energy emission, in
colliding-wind binaries.

With the advent of the upgrade to the Very Large Array, the EVLA, it
will be possible to attain radiometry at least an order of magnitude
better than currently available. Further, it will be possible to
observe the entire continuum spectrum from 1 to 50 GHz with
unprecedented precision. The Atacama Large Millimetre Array (ALMA) is
another exciting development that will have the ability to image dust
emission structures around massive stars, most particularly older,
cooler dust that exists further from the underlying WC binary systems.

The ability to readily identify binaries through the radio, X-ray and
IR characteristics of CWBs gives a very powerful tool for extending
our understanding of the birth-rate and formation environments of
massive stellar binaries. This is especially timely given the recent
discoveries of more massive, optically obscured stellar clusters in the
Milky Way, especially in the direction of the Galactic Centre.

\acknowledgements Much of the work described in this brief review was
done in collaboration with many colleagues. Most particularly, I have
had many engaging discussions with both Julian Pittard (Leeds
University) and Perry Williams (IfA, Edinburgh) as part of a number of
successful projects related to colliding-winds. Also I like to thank a
number of very enthusiastic students who have worked with me on
various CWB projects over the past six years - Nick Bolingbroke, Evan
O'Connor, Marshall Kennedy and Thomas Johnson. Lastly, I'm grateful to
Gemma Anderson for sharing some of the results from the ChIcAGO
project prior to publication.


\end{document}